\documentstyle[preprint,aps,epsf,axodraw,tighten]{revtex}

\begin{document}

\newcommand{\dfrac}[2]{\frac{\displaystyle #1}{\displaystyle #2}}
\draft
\preprint{VPI--IPPAP--00--02}

\title{Universal Torsion--Induced Interaction from Large Extra
Dimensions}
\author{
Lay~Nam~Chang${}^{(1)}$\thanks{electronic address: laynam@vt.edu},
Oleg~Lebedev${}^{(1)}$\thanks{electronic address: lebedev@quasar.phys.vt.edu},
Will~Loinaz${}^{(1,2)}$\thanks{electronic address: loinaz@alumni.princeton.edu}, and
Tatsu~Takeuchi${}^{(1)}$\thanks{electronic address: takeuchi@vt.edu}
}
\address{${}^{(1)}$Institute for Particle Physics and Astrophysics,
Physics Department, Virginia Tech, Blacksburg, VA 24061\\
${}^{(2)}$Department of Physics, Amherst College, Amherst MA 01002
}

\date{\today}
\maketitle

\begin{abstract}
We consider the Kaluza--Klein (KK) scenario in which only gravity exists in 
the bulk. Without the {\em assumption} of symmetric connection, 
the presence of brane fermions induces torsion.   
The result is a universal axial contact interaction that dominates those 
induced by KK gravitons. This enhancement arises from a large spin 
density on the brane.
Using a global fit to $Z$--pole observables, we find the 3$\sigma$ bound on 
the scale of quantum gravity to be 28 TeV for $n=2$. 
If Dirac or light sterile neutrinos are present, 
the data from SN1987A increase the bound to 
$\sqrt{n}M_S \geq 210\; {\rm TeV}$.
\end{abstract}


\narrowtext

Consistent string theories require dimensions beyond the usual four, 
that could in principle be as large as a millimeter.
These large extra dimensions provide new avenues for 
solution to the hierarchy problem \cite{Arkani-Hamed:1998rs}
and have significant phenomenological consequences as well.
Current collider experiments set a lower bound of about 1 TeV on 
the fundamental Planck scale, while cosmological considerations 
require it to be around 50 TeV for two extra dimensions.  
Future colliders will be able to probe up to 5$\sim$8 TeV \cite{junk}.

This paper considers the minimal Kaluza--Klein scenario (KK) wherein
only gravity exists in the bulk while all Standard Model fields are
localized on a 4D brane. 
Without {\it a priori} assumptions on its symmetry, 
fermions always induce antisymmetric pieces, or torsion,
in the gravity connection.
We show that this implies 
a universal $U(45)$--invariant contact interaction which is
suppressed only by the square of the {\it fundamental} Planck scale. 
This interaction therefore dominates four--fermion interactions induced by
KK graviton exchange at current collider energies 
and provides important information concerning the viability of extra
dimensions.
The enhancement is distinct from the KK mechanism.
It originates
from a large $4+n$ dimensional spin density due to the spinor fields
confined to a 4 dimensional brane.
We obtain a limit on this interaction from
electroweak precision data and compare it to other constraints available
from particle physics and astrophysics.

While the metric is coupled to the energy--momentum tensor in gravity,
torsion is coupled directly to the spin density of matter 
systems\cite{Kibble:1961ba,Hehl:1976kj}.  It appears in any
description of gravity where Lorentz transformations are treated
as local symmetries and is a feature of string theory and other variations 
of general relativity.
The basic quantity is the torsion tensor $T^{\alpha}{}_{\beta\gamma}$,
defined as the antisymmetric part of the connection
$\tilde{\Gamma}^{\alpha}{}_{\beta\gamma}$:
\begin{equation}
T^{\alpha}{}_{\beta\gamma} = 
\tilde{\Gamma}^{\alpha}{}_{\beta\gamma}-
\tilde{\Gamma}^{\alpha}{}_{\gamma\beta}.
\end{equation}
Torsion violates the equivalence principle 
in its
{\it very strong} form \cite{wheeler}, as it cannot be removed by an
appropriate choice of coordinates. It does not
directly affect the propagation of light and test particles, and
thus cannot be probed by standard tests of general relativity
\cite{wheeler}.

We denote with a tilde quantities derived from
a non--symmetric connection $\tilde{\Gamma}^{\alpha}{}_{\beta\gamma}$,
while those without a tilde refer to quantities
derived from Christoffel symbols
$\Gamma^{\alpha}{}_{\beta\gamma}$.\footnote{%
We follow the conventions and definitions of Ref.~\cite{buch} except
for the definition of $\gamma_5$.}
Via the metric condition $\tilde{\nabla}_{\alpha} g_{\mu\nu}=0$, 
the torsion tensor can be related to the contorsion tensor
$K^{\alpha}{}_{\beta\gamma}$ defined by
\begin{equation}
\tilde{\Gamma}^{\alpha}{}_{\beta\gamma}=
\Gamma^{\alpha}{}_{\beta\gamma} + K^{\alpha}{}_{\beta\gamma},
\end{equation}
such that 
$K_{\alpha\beta\gamma} = \frac{1}{2}\left( T_{\alpha\beta\gamma}
-T_{\beta\alpha\gamma}-T_{\gamma\alpha\beta} \right)$.
${T^{\alpha}}_{\beta\gamma}$ in 4D contains 24 independent components.
However, only its totally antisymmetric part, expressible as an axial
vector $S^\sigma$, 
\begin{equation}
S^\sigma = i\,\epsilon^{\mu\nu\rho\sigma} T_{\mu\nu\rho}
\label{pseudovector}
\end{equation}
is relevant for spin 1/2 fermions \cite{buch}.
In what follows we assume that torsion is completely antisymmetric.
The relations above (except for Eq.~(\ref{pseudovector})) can be
generalized for the case of $4+n$ dimensions in a straightforward
manner.
We will use the following convention for the indices \cite{Han:1999sg}:
$\mu=1,\cdots,4$, $\hat{\mu}=1,\cdots,4+n$, $i=5,\cdots,4+n$. 
The signature of the metric is $(1,-1,\cdots,-1)$.

To begin, note that torsion minimally couples to fermions only \cite{buch}.
The minimal action for $4+n$ dimensional gravity coupled to fermions
localized on a 4--dimensional brane is: 
\begin{eqnarray}
\lefteqn{
S = -\frac{1}{\hat{\kappa}^2} \int d^{4+n}x \;
     \sqrt{\vert \hat{g}_{4+n}\vert} \;\tilde{R}
} & &  \label{action}
\\
& + & \int d^{4}x \;\sqrt{\vert \hat{g}_4 \vert}\; \frac{i}{2}
         \left[ \bar\Psi\gamma^\mu\tilde\nabla_\mu\Psi
              - \left( \tilde\nabla_\mu\bar\Psi \right)\gamma^\mu\Psi
              + 2 i M \bar\Psi \Psi
         \right],
\nonumber	
\end{eqnarray}	
Here $\hat{\kappa}^2=16\pi G_N^{(4+n)}$, $\tilde{R}$ is the $4+n$
dimensional scalar curvature, and $\hat{g}_{4+n}$ and $\hat{g}_4$
are respectively the $4+n$ and $4$--dimensional (induced) metric
determinants. 

The covariant derivative $\tilde\nabla_{\mu}$ is defined by 
$\tilde\nabla_{\mu}\Psi=\partial_{\mu}\Psi
 +\frac{i}{2}\tilde\omega^{ab}_{\mu}\sigma_{ab}\Psi$,  where
$\tilde\omega^{ab}_{\mu}$ is the spin--connection,
$\sigma_{ab}=\frac{i}{2}\left[ \gamma_a, \gamma_b \right]$, 
with $a$, $b$  the local Lorentz indices. 
A general spin connection $\tilde\omega^{ab}_{\mu}$ can be expressed
in terms  of a torsion-free spin-connection
$\omega^{ab}_{\mu}$, the contorsion tensor and the vierbein $e^a_\mu$:
\begin{equation}
\tilde\omega^{ab}_{\mu} = \omega^{ab}_{\mu}
+ \frac{1}{4} K^{\nu}{}_{\lambda\mu}
  \left( e^{\lambda a} e^b_\nu - e^{\lambda b} e^a_\nu \right).
\end{equation}
Upon substitution, the action in the case of completely antisymmetric
torsion becomes
\begin{eqnarray}
\lefteqn{
S = -\frac{1}{\hat{\kappa}^2} \int d^{4+n}x \;\sqrt{\vert \hat{g}_{4+n} \vert}
    \;\left( R - K^{\hat{\mu}\hat{\nu}\hat{\rho}}
                 K_{\hat{\mu}\hat{\nu}\hat{\rho}}
      \right)
} \cr
& + & \int d^{4}x \;\sqrt{\vert \hat{g}_4 \vert}\;i
          \bar\Psi
          \left( \gamma^{\mu}\nabla_{\mu}
               - \frac{1}{8}S^\mu \gamma_\mu \gamma_5
               + iM
          \right) \Psi\;.
\label{action1}
\end{eqnarray}
Here $R$ is the $4+n$ dimensional {\it metric} curvature,
$K^{\hat{\mu}\hat{\nu}\hat{\rho}} = 
\frac{1}{2}\;T^{\hat{\mu}\hat{\nu}\hat{\rho}}$, and
$\nabla_{\mu}$ is the conventional covariant derivative
without torsion.  We use
$\gamma_5=-i\gamma^0\gamma^1\gamma^2\gamma^3$.
The resultant equations of motion are
\begin{eqnarray}
&& K_{i\hat{\mu}\hat{\nu}} = 0\;,\cr
&& S_{\mu} = i\frac{3}{2} 
               \frac{ \sqrt{\vert \hat{g}_4 \vert }  }{ \sqrt{ \vert \hat{g}_{4+n} \vert} }\;
               \hat{\kappa}^2\;
               \bar\Psi\gamma_{\mu}\gamma_5\Psi\;
               \delta^{(n)}(x)\;, \cr
&& R_{\hat{\mu}\hat{\nu}} - \frac{1}{2} g_{\hat{\mu}\hat{\nu}} R
     = \frac{ \hat{\kappa}^2 }{ 2 }\;T_{\hat{\mu}\hat{\nu}}
     + {\cal O}\left( \hat{\kappa}^4 \right) \;.
\label{eqmotion}
\end{eqnarray}
$T_{\hat{\mu}\hat{\nu}}$ is the torsion--free
energy--momentum tensor for the matter fields.
The first two of these relations are algebraic constraints.
Classically, torsion does not propagate
and is zero outside the matter distribution
\cite{Hehl:1976kj}.
Its source is the {\it spin density} of fermions confined on the brane.
Elimination of $S_\mu$ from the action via Eq.~(\ref{eqmotion})
produces a fermion contact interaction:
\widetext
\begin{eqnarray}
S & = & -\frac{1}{\hat{\kappa}^2} \int d^{4+n}x
\;\sqrt{\vert \hat{g}_{4+n} \vert}\;R
        \\  \label{action2}
  &   & + \int d^{4}x \;\sqrt{\vert \hat{g}_4 \vert}\;
        \left[ \bar\Psi\left( i\gamma^{\mu}\nabla_{\mu} - M \right)\Psi
             + \frac{3}{32} 
               \frac{ \sqrt{ \vert \hat{g}_4 \vert } }{ \sqrt{\vert \hat{g}_{4+n} \vert} } 
               \hat{\kappa}^2\,
               \left( \bar\Psi\gamma_{\mu}\gamma_5\Psi \right)^2\,
               \delta^{(n)}(0) 
        \right]\;. \nonumber
\end{eqnarray}
\narrowtext
The delta--function appearing in this expression should be regularized
to account for a finite brane width:
\begin{eqnarray}
\delta^{(n)}(0) \rightarrow \frac{1}{(2\pi)^n} \int_{0}^{M_S} d^n k =
\dfrac{ M_S^n }{ 2^{n-1} \pi^{n/2}\,n\,\Gamma\!\left(\dfrac{n}{2}\right) }\;,
\end{eqnarray}
$M_S$ is the cutoff scale of the effective theory, here taken
to be of the order of the inverse brane width. The $4+n$
dimensional coupling constant $\hat{\kappa}$ is related to the 4--dimensional
coupling $\kappa$ and the volume of the extra dimensions compactified on a 
torus via $\hat{\kappa}^2=
\kappa^2 V_n=16\pi(4\pi)^{n/2}\Gamma(n/2)M_S^{-(n+2)}$
\cite{Han:1999sg}\footnote{For simplicity we set the string scale
and the $4+n$ dimensional Planck mass equal.}.
As a result, the
leading ${\cal O}\left( \hat{\kappa}^2 \right)$ torsion contribution
to the action is given by
\begin{equation}
\Delta S
= \int d^4x\;\frac{3\pi}{n M_S^2}
      \left[ \sum_j \bar\Psi_j\gamma_{\mu}\gamma_5\Psi_j
      \right]^2\;,
\label{action3}
\end{equation}
where $j$ runs over all fermions existing on the brane.
The expansion in $\hat{\kappa}$ is expected to be valid provided the
typical energy $E$
of a physical process is  below the cutoff scale $M_S$
(see also Ref.~\cite{Giudice:1999ck}). 
In this case a typical ``size'' of torsion is considerably below $E$.
Note that the exact coefficient in Eq.(\ref{action3}) depends on the 
assumptions about the short-distance physics; in particular, it 
depends on the regularization of the delta-function.

The interaction (\ref{action3}) is the unique contact interaction
possessing the maximal approximate global symmetry of the minimal Standard
Model, {\it i.e.} the group $U(45)$
acting on  the 45 Weyl spinors
$\Psi_L=(q_L, u^c_R,d^c_R,l_L,e^c_R)$
\cite{Buchmuller:1997hn}\footnote{Other
contact interactions
possessing the same symmetry can be brought into the form of
Eq.~(\ref{action3}).}.
The effect is truly universal for all fermions, in contrast to the
four--fermion operators induced by KK graviton exchanges.
Graviton couplings
are mass and energy dependent, leading to different strengths 
for different fields. Finally, the KK--induced interactions 
have two additional suppression factors:
$s/M_S^2$ \cite{Han:1999sg} and  $f^2/M_S^2$ \cite{Bando:1999di}. 
The former follows
from the graviton coupling to the energy--momentum tensor, the latter
from the brane recoil effects (note that the rigidity of the
brane $f$ plays no role in our argument).
Consequently, at typical accelerator energies
the interaction (\ref{action3}) is enhanced over the KK--induced 
interactions by {\it orders of magnitude} ({\it e.g.} at LEP energies
the enhancement factor is about $10^2$) and will completely dominate.
This enhancement results from a large $4+n$ dimensional 
{\it spin density} on the brane and is present whenever fermions are 
localized. Note that the interaction (\ref{action3}) is  repulsive
for aligned spins \cite{Hehl:1976kj}.


Let us briefly discuss how this result is modified by quantum 
corrections \cite{cs}.
Generally, fermion loops will induce propagation of torsion along the brane, 
with $S_\mu \bar\Psi\gamma^{\mu}\gamma_5\Psi$ generating the relevant
kinetic terms.
The associated quantum of torsion will have a mass of order $M_S$.  
Its propagation effects are therefore
irrelevant at typical accelerator energies.  
Loop corrections to the tree level torsion coupling
and mass evaluated using an explicit cutoff amount to a rescaling
of $M_S$ in Eq.~(\ref{action3}) by a factor of 1$\sim$2. 
However, if we use dimensional regularization,
the result is largely insensitive to radiative corrections
due to the absence of quadratic divergences.

The universal interaction (\ref{action3}) will affect
$Z$--pole electroweak observables.  Corrections to the oblique parameters
appear at the two--loop level and can be neglected
{\footnote{Throughout this analysis we retain only leading
tree or one--loop contributions.  This approximation is valid
to leading order in $1/n$.}}. 
Two types of vertex corrections are shown in Fig.~1. 
The combined contribution of the diagram
in Fig.~1a and the corresponding wave function renormalization diagrams  
is suppressed by $1/M_S^2$ and leads only to a {\it universal} 
multiplicative correction to the couplings (neglecting light fermion
masses).  
Since the observables we will consider are ratios of couplings, such
corrections will cancel.
The diagram in Fig.~1b is significant.  Note that
the corresponding wave function renormalization diagram vanishes.
Summing contributions of all of the fermions and taking into account
$\sum I_3=0$ , we write the leading correction to the
$Z$--couplings as
\begin{equation}
\delta h_L=-\delta h_R={3N_c m_t^2 \over 4\pi n M_S^2}\;
\label{delta}
\ln {M_S^2 \over m_t^2}\;,
\end{equation}
where the Z--vertex is defined as $-i{g\over \cos \theta_W}
Z_\mu \bar\Psi \gamma^\mu (P_Lh_L+P_Rh_R)\Psi$.  The contribution of
torsion to $\delta h_L$ is strictly positive.

We perform a global fit to the LEP/SLD electroweak observables including
$R_{\nu/\ell} = \Gamma(Z\rightarrow \nu\bar{\nu})/
\Gamma(Z\rightarrow \ell^+\ell^-)$, $R_{b,c}$,
$A_{FB}(i)$ and $A_i$ ($i=e,\mu,\tau,b,c$)
{\footnote{We omit $R_\ell$ from our fit since a universal correction to 
$R_\ell$ only shifts the value of $\alpha_s(M_Z)$.}}.
The data used and a detailed description of the technique (applied to a
different model) can be found in \cite{Lebedev:2000vc}.
Note that the KK graviton radiative corrections
are suppressed as discussed above;
moreover, the KK vertex corrections will largely cancel in our set
of observables since, for the case of light fermions,
they modify the couplings multiplicatively.
The KK graviton corrections to the
oblique parameters may not be negligible \cite{oblique}
and can affect our fit results through $\sin^2\theta_W$.
In the fit, we leave
$\delta s^2 \equiv \sin^2\theta_W - [\sin^2\theta_W]_{\rm SM}$ 
as a free parameter to account for a
variation in the Higgs mass and KK graviton radiative
corrections. From a two--parameter fit we obtain
\begin{eqnarray}
\delta h_L & = & -0.00049 \pm 0.00021 \cr
\delta s^2 & = &  -0.00068 \pm 0.00018 \; .
\end{eqnarray}
The $\chi^2/d.o.f.$ of the fit is 17/12.
Fig.~2 shows that $\delta h_L$ is most strongly constrained by 
$R_{\nu/\ell}$.
Since the experimental value for $R_{\nu/\ell}$
is about $2\sigma$ below the SM prediction, the preferred value
of $\delta h_L$ is about $2\sigma$ below zero.  
The value of $\delta h_L$ is almost uncorrelated 
with $\delta s^2$ and thus with the Higgs mass.  As a 
result of this classical statistical analysis, 
the model is excluded at the $2\sigma$ level since it generates
only a positive $\delta h_L$.  Using Eq.~(\ref{delta}), we obtain the 
$3\sigma$ bound on $M_S$:
\begin{equation}
M_S \geq 28 \; {\rm TeV} 
\end{equation}
for $n=2$.  For $n=4 (6)$ the bound weakens to 19 (15) TeV.
This implies that we do not expect deviations from Newton's law
at distances above $9 \times 10^{-4}$ mm for $n=2$.

We next consider other constraints on the universal
interaction{\footnote{Phenomenological implications of 4--d gravity
with torsion were also considered in \cite{belyaev}.}}.
$A\times A^{+}$ contact interaction (\ref{action3}) affects
at the tree level
the differential cross sections for $e^+e^- \rightarrow f\bar f$
measured at LEP.
The OPAL measurements \cite{opal} imply
\begin{equation}
\sqrt{n} M_S \geq 10.3\;{\rm TeV}
\end{equation}
at the 95\% confidence level. Electron--quark contact interactions can
also be constrained via HERA DIS data, Drell--Yan production at the
Tevatron, {\it etc.}
The global analysis \cite{cheung} yields
\begin{equation}
\sqrt{n} M_S \geq 5.3\;{\rm TeV}\;.
\end{equation}
Another potentially strong constraint can come from the measurement
of the invisible width of the $\Upsilon$ and $J/\Psi$ resonances
at B and $\tau$--charm factories \cite{Chang:1998tq}.

A powerful astrophysical constraint can be derived if we admit 
existence of Dirac or light sterile neutrinos.
For the case of Dirac neutrinos, the torsion--induced interaction
containes a term
\begin{equation}
\Delta {\cal{L}}=
-\frac{6\pi}{n M_S^2}\;
\bar q\gamma^{\mu}\gamma_5 q \; \bar \nu_R\gamma_{\mu} \nu_R\;.
\end{equation}
This contact interaction provides a new channel of energy drain
during neutron star collapse, since right handed neutrinos
produced by nucleon interactions leave the core without rescattering
\cite{grifols}. This would affect neutron star
evolution; in particular, it would modify the duration of the standard
neutrino burst. From observations of SN 1987A one infers \cite{grifols}
\begin{equation}
\sqrt{n} M_S \geq 210\;{\rm TeV}\;.
\label{SN}
\end{equation}
Similar considerations apply to the case of light sterile neutrinos:
if $m_{\nu_s} \ll 50\;{\rm MeV}$, the core temperature, the analysis is
completely analogous to that of Dirac neutrinos and the bound
(\ref{SN}) holds.  This bound translates into an upper bound 
on the compactification radius of $3 \times 10^{-5}$ mm for $n=2$.

Since $M_S$ controls all gravity effects in extra dimensions, the limits
on $M_S$ being larger than tens of TeV
reported here imply weaker KK graviton couplings than those considered in
the literature.  The limits were obtained under a minimal set of assumptions
in the context of physics of extra dimensions.
We first consider the  general set of connections consistent 
with general covariance and local Lorentz symmetries.
We fix all matter fields to be on a brane, consistent with the most
conservative scenario.  
The consequence is that tree--level effects from a minimal
action are sufficiently strong to produce the bounds reported above.
Even for less conservative scenarios
the universal interaction (\ref{action3})
can be expected to dominate possible KK gauge effects
as long as the fermions are confined to the brane.
We emphasize that this interaction is generic unless we impose the
additional condition that the connection be symmetric.\footnote{%
The analogous four--fermion interaction involving gauginos in
supergravity models needs a separate discussion because of their 
connections to the corresponding coupling among gauge bosons.  
However these terms do not affect the kinds of phenomenology
addressed in the present paper.}
Finally, it is interesting to note that, in this context, particle physics
places a constraint on violation of the strong equivalence principle.

This research is supported in part by a grant from the U.S. 
Department of Energy, DE--FG05--92ER40709.


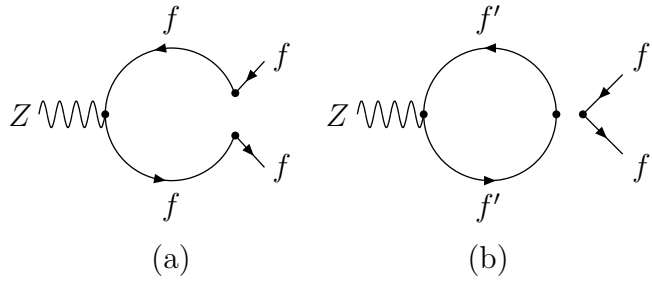
\begin{figure}[h]
\begin{center} 
\begin{picture}(300,100)(10,-10)
\Vertex(50,50){1.5}
\Vertex(99,58){1.5}
\Vertex(99,42){1.5}
\Photon(25,50)(50,50){5}{4}
\ArrowArc(75,50)(25,20,180)
\ArrowArc(75,50)(25,180,340)
\ArrowLine(110,70)(99,58)
\ArrowLine(99,42)(110,30)
\Text(18,50)[]{$Z$}
\Text(75,86)[]{$f$}
\Text(75,15)[]{$f$}
\Text(117,72)[]{$f$}
\Text(117,30)[]{$f$}
\Vertex(170,50){1.5}
\Vertex(220,50){1.5}
\Vertex(230,50){1.5}
\Photon(145,50)(170,50){5}{4}
\ArrowArc(195,50)(25,0,180)
\ArrowArc(195,50)(25,180,360)
\ArrowLine(245,65)(230,50)
\ArrowLine(230,50)(245,35)
\Text(138,50)[]{$Z$}
\Text(195,86)[]{$f'$}
\Text(195,15)[]{$f'$}
\Text(253,72)[]{$f$}
\Text(253,30)[]{$f$}
\Text(75,-5)[]{(a)}
\Text(195,-5)[]{(b)}
\end{picture}
\end{center}
\caption{Corrections to the $Zf\bar{f}$ coupling from the universal 
contact interaction Eq.~(\ref{action3}).} 
\label{Feynman}
\end{figure}


\begin{center}
\begin{figure}
\begin{picture}(240,220)(0,0)
\unitlength=1mm
\put(47,4){$\delta s^2$}
\put(0,42){$\delta h_L$}
\put(58,31.5){$R_{\nu/\ell}$}
\put(69,25){$R_b$}
\put(28,58){$A_{\rm LR}$}
\put(58,57){$A_{\rm FB}(b)$}
\epsfbox[0 0 240 220]{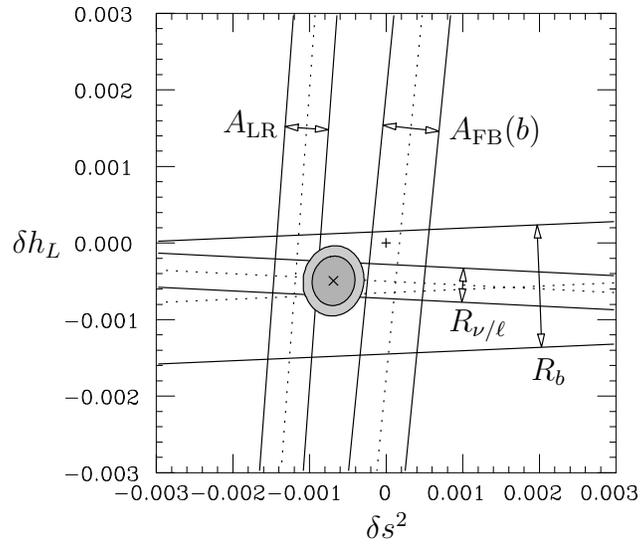}
\end{picture}
\caption{The 68\% and 90\% confidence contours in the
$\delta s^2$--$\delta h_L$ plane.
The $1\sigma$ bounds from the observables leading to the
strongest constraints are also shown.
}
\label{Fit}
\end{figure}
\end{center}

\narrowtext

\end{document}